\documentclass[sigconf]{acmart}
\pdfoutput=1
\renewcommand\footnotetextcopyrightpermission[1]{} %
\makeatletter
\renewcommand\@formatdoi[1]{\ignorespaces}
\makeatother
\PassOptionsToPackage{bookmarks=false}{hyperref}
\usepackage{todonotes}
\usepackage{algorithm}
\usepackage[noend]{algpseudocode}
\usepackage{enumitem}
\usepackage{tabularx}

\makeatletter \let\c@table\c@figure \makeatother

\graphicspath{{../}}

\hypersetup{draft}
\begin{document}
\title{BioGen: Automated Biography Generation}
\author{Heer Ambavi*, Ayush Garg*, Ayush Garg*, Nitiksha*, Mridul Sharma*, Rohit Sharma* \and Jayesh Choudhari, Mayank Singh}
\affiliation{Department of Computer Science and Engineering, Indian Institute of Technology Gandhinagar, India}

\email{singh.mayank@iitgn.ac.in}


\authornote{Equal contribution}

\begin{abstract}  
A biography of a person is the detailed
description of several life events including his education, work, relationships and death. Wikipedia, the free web-based encyclopedia, consists of millions of manually curated biographies of eminent politicians, film and sports personalities, etc.   However, manual curation efforts, even though efficient, suffers from significant delays.  In this work, we propose an automatic biography generation framework \texttt{BioGen}.  \texttt{BioGen} generates a short collection of biographical sentences clustered into multiple events of life.  Evaluation results show that biographies generated by \texttt{BioGen}  are significantly closer to manually written biographies in Wikipedia. A working model of this framework is available at  \textit{\url{nlpbiogen.herokuapp.com/home/}}
\end{abstract}

\maketitle

\section{Introduction}
\label{sec:intro}
As Internet technology continues to thrive, a large number of documents are continuously generated and published online. Online newspapers, for instance, publish articles describing important facts or life events of well-known personalities.  However, these documents being highly unstructured and noisy, contain both meaningful biographical facts, as well as unrelated information to describe the person, for example, opinions, discussions, etc. Thus, extracting meaningful biographical sentences from a large pool of unstructured text is a challenging problem. Although humans manage to filter the desired information, manual inspection does not scale to very large document collections. 

\subsection{Textual Biographies}
A textual biography can be represented as a series of facts and events that make up a person's life. Different types of biographical facts include aspects related to personal characteristics like date and place of birth and death, education, career, occupation, affiliation, and relationships.  Overall, a general biography generation process can be described in three major steps: (i) identifying biographical sentences, (ii) classifying biographical sentences into different life-events, and (iii) relevancy-based ranking of sentences in each life-event class. Along with the biographical information, a Wikipedia profile of a person also consists of a consistently-formatted table on the top right-hand side of the page. This table or box is known as an \emph{infobox}, and it contains some important facts related to the person.

\subsection{Machine Learning in Biography Generation}
Majority of past literature focuses on Machine Learning techniques for information extraction. Zhou et al.~\cite{zhou2005multi} trained biographical and non-biographical sentence classifiers to categorize sentences. They also employ a Naive Bayes model with n-gram features to classify sentences into ten classes such as bio, fame factor, personality, etc. This work looks similar to ours, but this requires a good amount of human effort. Biadsy et al.~\cite{biadsy2008unsupervised} proposed summarization techniques to extract important information from multiple sentences. 
Liu et al.~\cite{liu2018generating} also use multi-document summarization. For identifying salient information, the paragraphs are ranked and ordered using various extractive summarization techniques. However, both these systems (\cite{biadsy2008unsupervised} and \cite{liu2018generating}) do not focus on sectionizing the biography.
The works by Filatova et al.~\cite{Filatova:2005} and Barzilay et al.~\cite{Barzilay:2001} focus on specific tasks such as identifying occupation related important events and sentence ordering techniques respectively.
One of the recent works in generating sentences is by Lebret et al.~\cite{lebret2016neural}. They use \emph{concept to text generation} approach to generate only single/first sentence using the fact tables present in Wikipedia.

\subsection{Our Contribution}
In this paper, we address the task of automatically extracting biographical facts from textual documents published on the Web. We pose this problem in the extractive summarization framework and propose a two-stage extractive strategy. In the first stage, sentences are classified into biographical facts or not. In the following stage, we classify biographical sentences into several life-event categories. Along with the biography generation task, we also propose a method to generate \emph{Infobox} which is a consistently-formatted table mentioning some important facts and events related to a person. We experimented with several ML models and achieve significantly high F-scores. 

\noindent \textbf{Outline:} Section 2 describes the datasets that are used to train the models. Section 3 describes the components of our system in more detail. Section 4 describes our experiments and results. Section 5 draws the final conclusion of our work. 

\section{Datasets}
\label{sec:datasets}
The current work requires large textual biography datasets. Also, in order to categorize between biographical and non-biographical sentences, we leverage a non-biographical news dataset. Following is the descriptions of the dataset used.

{\texttt{TREC-RCV1}~\cite{lebret2016neural}: }
This Reuters news corpus consists of $\sim$8.5 million news titles, links and timestamps collected between Jan 2007 and Aug 2016. Dataset was used for training a 2-class classifier in the first step of the biography generation process (see Section ~\ref{sec:2_class}). All the sentences in this dataset are labeled as \emph{non-biographical}.

{\texttt{WikiBio}~\cite{Amini:2009}:}
This dataset consists of $\sim$730K biographical pages from English Wikipedia. For each article, the dataset consists of only the first paragraph. This dataset was used for training the 2-class classifier as mentioned in Section \ref{sec:2_class}. All the sentences are labeled as \emph{biographical} sentences. 

{\texttt{BigWikiBio}:}
We curated this dataset by crawling English Wikipedia articles. It consists of $\sim$6M Wikipedia biographies. This dataset was used to train the 6-class classifier (see Section \ref{sec:6_class}). 

\section{Methodology}
 The biography generation process involves multi-stage extractive subtasks. In this section, we describe these stages in detail. Along with the biographical information, a Wikipedia page also consists of an \emph{infobox}. For the sake of completeness, we also describe an automatic approach to generate infoboxes similar to the one present on the Wikipedia page.
 
 \subsection{Identifying Biographical Sentences}
\label{sec:2_class}
A textual document that describes an event or some news related to a person contains a large number of non-biographical sentences  as compared to biographical sentences. In the first stage, sentences were categorized into the above two categories.  

\subsubsection{Data Pre-processing} Given a text document, we partition it into a set of sentences. Next, we enrich extracted sentences by performing standard NLP tasks like special character removal, spell check, etc. 

\subsubsection{Sentence representation and classification}
Each sentence is converted into a fixed-length \texttt{TF-IDF} vector representation. We consider sentences available in the \texttt{TREC} dataset as \textit{non-biographical} sentences. Whereas sentences in \texttt{WikiBio} dataset are considered as \textit{biographical}. We experiment with several machine learning models like Logistic Regression, Decision Trees, Naive Bayes, etc., to perform binary classification.  Since the Logistic Regression model performed best (evaluation scores described in Section~\ref{sec:results}), we leverage its classified results for next stages.

\subsubsection{Filtering False Positives}
Our classifier resulted in some false positives --- sentences that are non-biographical but were classified as biographical. We, further, filter out these cases by employing standard \textit{Named Entity Recognition} technique. We consider only those sentences that contain at least one of the three named entities: (i) \emph{PERSON}, (ii) \emph{PLACE} or (iii) \emph{ORGANIZATION}. In the next stage, we classify biographical sentences into several life-event categories.

\subsection{Classifying Biographical Sentences}
\label{sec:6_class}
The categorized biographical sentences are further classified into  six life-event classes namely \emph{Education, Career, Life, Awards, Special Notes, and Death}. We leverage section information in \texttt{BigWikiBio} dataset (described in Section \ref{sec:datasets}) to construct a mapping between sentences and life-events. We label each sentence in the Wikipedia page with its corresponding section heading and further map it to a broad life-event class. For example, sentences with sections headings as \emph{College, High School, Early life and education, Education, etc.} are labeled as \emph{Education} class, \emph{Politics, Music career, Career, Works, Publications, Research, etc.} are labeled as \emph{Career} class, \emph{Honors, Awards, Recognition, Championships, Achievements, Accomplishments, etc.} are labeled as \emph{Honours/Awards} class, \emph{Honors, Awards, Recognition, Championships, Achievements, Accomplishments, etc.} are labeled as \emph{Honours/Awards} class, \emph{Notes, Legacy, Personal, Gallery, Influences, Other, Controversies, etc.} are labeled as \emph{Special Notes} class, and \emph{Death, Death and Legacy, Later life, and Death, etc.} are labeled as \emph{Death} class.

Next, we leverage the Logistic Regression model to perform this multi-class classification task. We construct similar fixed-length \texttt{TF-IDF} vector representation as described in Section~\ref{sec:2_class}. The classification results into clusters of similar sentences representing a single life-event of person.

\subsection{Summarization}
\label{sec:summarizer}
A single life-event cluster might contain hundreds of biographical sentences. We, therefore, rank the most important sentences by leveraging graph ranking algorithm\footnote{ We use Gensim implementation~\cite{rehurek_lrec}.} \textit{Text Rank}~\cite{mihalcea2004textrank}. For a given person, we apply \textit{Text Rank} on each of the obtained six clusters. The ranking imparts flexibility in experimenting with multiple length values of the generated biography.

\subsection{Generating Infobox}
\label{sec:infobox}
The infobox is a well-formatted table which gives a short and concise description of important facts related to the person. We use the following set facts in our proposed \textit{Infobox} of a queried person.
\begin{itemize}
    \item \textbf{Name:} Name of the queried person.
    \item \textbf{Date of Birth \& Date of Death:} We use regular expressions to extract the date, depending on the context phrases such as `born on', `birth', etc. 
    \item \textbf{Place of birth:} We use a similar methodology as above. We leverage part-of-speech (POS) tagging and Named Entity Recognition\footnote{We use the Stanford NER library\cite{Bird:2009:NLP:1717171}.} to identify the place of the birth.
    \item \textbf{Awards:} We extract award information by leveraging a list of all the awards available at: \emph{Wikipedia List of Awards page} \footnote{https://en.wikipedia.org/wiki/List\_of\_awards}. We, next, use standard string matching to identify an award name in the biographical sentences. 
    \item \textbf{Education \& Career:} Here also, we leverage education information (degree, courses, etc.) present at official government sites like \texttt{data.gov.in} and \texttt{usa.gov}.  Similarly, career-related information was obtained using the Wikipedia page \emph{list of Occupations}.\footnote{https://en.wikipedia.org/wiki/Lists\_of\_occupations}
\end{itemize}
As an additional feature, we also present a profile image of the person by performing an image-based Google search query. This additional feature enriches textual biography with visual aspects similar to a Wikipedia profile of a person. 
\section{Experiments and Results}
\label{sec:results}
In this section, we present evaluation results of two sub-tasks: (i) \emph{Biography Generation} and (ii) \emph{Infobox Generation}.

\subsection{Tasks and Evaluation Measures:}
The evaluation metrics for the above tasks are as follows:

\subsubsection{Biography Generation} We compare biographies generated by \texttt{BioGen} against corresponding Wikipedia page. Note that, biography generation task is similar to document summarization. We, therefore, use \emph{ROUGE} score to evaluate our generated biographies. In the current paper, we three variations of ROUGE, \emph{ROUGE-1}, \emph{ROUGE-2}, and \emph{ROUGE-L} scores.

\subsubsection{Infobox Generation}
To evaluate the quality of the infobox generated by \texttt{BioGen} we define a score for each field in the infobox. Let $I_{BG}$ and $I_{W}$ be the infobox generated by \texttt{BioGen} and the infobox present on Wikipedia page respectively. Let $f_{BG} \in I_{BG} = \{f_{BG}^{(c)}\}$ be the set of characteristics present in the field recovered by \texttt{BioGen} and $f_{W} \in I_{W} = \{f_{W}^{(c)}\}$ be the corresponding field set in the infobox present on the Wikipedia page. Then, the score for each field $f_{BG} \in I_{BG}$ is defined as:
    \begin{math}
        S(f_{BG}) = \frac{\sum_{c \in f_{BG}}[f_{BG}^{(c)} \in f_{W}]}{|f_{W}|}
    \end{math}
    And the total score for the generated infobox $I_{BG}$ is given by the average over the present fields given as:
    \begin{math}
        S(I_{BG}) = \frac{\sum_{f_{BG} \in I_{BG}} S(f_{BG})}{|I_{BG}|}
    \end{math}
    As the Infobox is generated containing the only specific set of fields which are Name, Date-Of-Birth,  Place-Of-Birth and Death, Awards, Education, and Career; the score is calculated only corresponding to those fields.

\subsection{Results}
\subsubsection{Biography Generation Accuracy}
Extracting information from an arbitrary webpage is a challenging task. We, therefore, leverage three web resources to construct a source document set for a given person query. The resources are \emph{Ducksters}\footnote{https://www.ducksters.com/}, \emph{IMDB}\footnote{https://www.imdb.com/}, and \emph{Zoomboola}\footnote{https://zoomboola.com/}. Ducksters is an educational site covering subjects such as history, science, geography, math, and biographies. IMDB is the world's most popular and authoritative source for movie, TV and celebrity content. Zoomboola is a news website. We experiment with randomly selected 150 biographies belonging to various domains such as Academics, Politics, Literature, Sports, Film Industry, etc.

Table \ref{tab:RS_1Source} describes the \emph{ROUGE} scores with the increasing number of sources that were used to generate the biography. We observe that \emph{ROUGE} score increases with an increase in the number of sources. I.e., the  similarity of biography generated by \texttt{BioGen} with that of the Wikipedia page increases as the number of sources increase. This also demonstrates the fact that a Wikipedia page is composed of multiple references.

\begin{table}[h]
    \centering
    \resizebox{\hsize}{!}{
    \begin{tabular}{|c | c c c| c c c| c c c|} 
     \hline
     &\multicolumn{3}{c|}{One Source}&\multicolumn{3}{c|}{Two Sources}&\multicolumn{3}{c|}{Three Sources}\\\cline{2-10}
     &F&P&R&F&P&R&F&P&R\\
     \hline
     Rouge-1 & 0.19 & 0.13 & 0.29& 0.22 & 0.15 & 0.31& 0.26 & 0.41 & 0.21\\ 
     \hline
     Rouge-2 & 0.05 & 0.03 & 0.10 & 0.06 & 0.04 & 0.10& 0.10 & 0.17 & 0.08 \\
     \hline
     Rouge-L & 0.16 & 0.15 & 0.33& 0.16 & 0.15 & 0.33 & 0.21 & 0.38 & 0.19 \\
     \hline
    \end{tabular}}
    \caption{Rouge Score (F1-Score, Precision, Recall) of generated text after summarization when the input document is taken from one, two and three sources.}
    \label{tab:RS_1Source}
    \vspace{-0.6cm}
\end{table}

Table \ref{tab:RS_WOSumm} shows the \emph{ROUGE} score when we do not use the \emph{Summarization} step of the biography generation process in \texttt{Biogen}. We can see that the Recall is better in this case, and this is because \emph{Summarization} step filters out some text from the biography, and thus the overlap with the Wikipedia page decreases.
\begin{table}[!tbh]
\centering
 \begin{tabular}{|c || c | c |} 
 \hline
 &  1-Source & 3-Sources \\ [0.5ex] 
 \hline
 Rouge-1  & 0.3869 & 0.5977 \\ 
 \hline
 Rouge-2 & 0.2071 & 0.2961 \\
 \hline
 Rouge-L & 0.3618 & 0.5698 \\[1ex] 
 \hline
\end{tabular}
\caption{Rouge scores (Recall values) without summarization with different number of sources.}
\vspace{-0.7cm}
\label{tab:RS_WOSumm}
\end{table}
Table \ref{tab:amitabhBio} shows Amitabh Bachchan's (Bollywood Film Industry Actor) biography generated using \texttt{BioGen} as an example. If we compare the generated biography to Amitabh Bachan's Wikipedia page\footnote{https://en.wikipedia.org/wiki/Amitabh\_Bachchan}, it turns out that the sentences which are highlighted in \emph{italics} are very much related to the class in which they have been classified, which says that \texttt{BioGen} does a fairly good job in not only generating relevant sentences but also placing them in the proper sections.
Also, we can see that the row corresponding to the field \emph{Death} is empty. I.e. \texttt{BioGen} did not extract any  information related to Amitabh Bachan's death, which is also true as per Wikipedia, as there is no information about his death. It is also important to note that \texttt{BioGen} did not add any arbitrary information in that field.

\begin{table*}[!tbh]
    \begin{tabularx}{\linewidth}{|l|X|}
    \hline
    \textbf{Career} & \textit{Bachchan's career moved into fifth gear after Ramesh Sippy's Sholay (1975).}The movies he made with Manmohan Desai (Amar Akbar Anthony, Naseeb, Mard) were immensely successful but towards the \textit{latter half of the 1980s his career went into a downspin.}However, the importance of being Amitabh Bachchan is not limited to his career, although he reinvented himself and experimented with his roles and acted in many successful films. \\
    \hline
    \textbf{Life} &  \textit{Amitabh Bachchan was born on October 11, 1942 in Allahabad. He is the son of late poet Harivansh Rai Bachchan and Teji Bachchan.Son of well known poet Harivansh Rai Bachchan and Teji Bachchan.He has a brother named Ajitabh.}He got his break in Bollywood after a letter of introduction from the then Prime Minister Mrs. Indira Gandhi, as he was a friend of her son, Rajiv Gandhi.\textit{He married Jaya Bhaduri, an accomplished actress in her own right, and they had two children, Shweta and Abhishek.}His son, Abhishek, is also an actor by his own rights. On November 16, 2011, he became a Dada (Paternal Grandfather) when Aishwarya gave birth to a daughter in Mumbai Hospital.\textit{He is already a Nana (maternal grandfather) to Navya Naveli and Agastye - Shweta's children.}After completing his education from Sherwood College, Nainital, and Kirori Mal College, Delhi University, he moved to Calcutta to work for shipping firm Shaw and Wallace.\\
    \hline
    \textbf{Awards} & \textit{In 1984, he was honored by the Indian government with the Padma Shri Award for his outstanding contribution to the Hindi film industry.France's highest civilian honour, the Knight of the Legion of Honour, was conferred upon him by the French Government in 2007, for his "exceptional career in the world of cinema and beyond"} \\
    \hline
    \textbf{Death} &  \\
    \hline
    \textbf{Rejected} &  Amitabh was in Goa during the last weekend to be one of the speaker at the THINK festival where he was honoured like he is being honoured in any and every place he makes his presence. At the very outset Bachchan was humble enough to let all those in the audience know that he held De Niro as one of his major sources of inspiration, was once forced to clear immigration in his hotel in Cairo, because his Egyptian fans became overly enthusiastic at the airport. Image caption\\
    \hline
\end{tabularx}
\caption{Amitabh Bachchan's biography generated by \texttt{BioGen}.}\label{tab:amitabhBio}
\vspace{-0.7cm}
\end{table*}

\begin{figure}[!tbh]
    \centering
    \includegraphics[width=1\hsize]{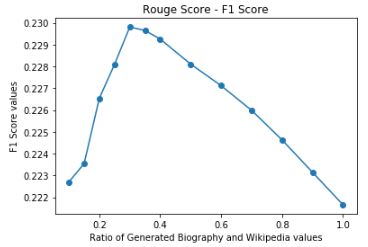}
    \caption{Change in \emph{ROUGE} score with changing ratio of lengths of \texttt{BioGen} generated and Wikipedia biographies.}
    \label{fig:F1Change}
\vspace{-0.3cm}
\end{figure}

We experimented \texttt{BioGen} with one more parameter, which is the length of the generated biography. Figure \ref{fig:F1Change} shows the change in the \emph{ROUGE} score with the change in the length of the biographies generated. Again, we can see that as the length increases the recall increases, but the precision decreases. Which is evident, because as we add in more and more content in the biography, we cover more and more information present on Wikipedia resulting in the increase of recall. However, as the length increases, the amount of new information that we get goes on reducing, which is seen by the decrease in the precision value.
Table~\ref{tab:amitabhBio} shows a sample biography generated for a famous Bollywood actor Amitabh Bachchan. It can be seen that our model does fairly well in extracting important sentences and classifies those into the six life-event classes. Also, as the actor is still alive, there should not be any `death' related event, which our model outputs correctly.
\subsubsection{Infobox Generation Accuracy}
Table \ref{tab:infobox_example} shows an example of an Infobox (for Amitabh Bachchan) generated using \texttt{BioGen}. This Infobox recovers almost all the information present on the original Wikipedia page\footnote{https://en.wikipedia.org/wiki/Amitabh\_Bachchan}, and achieves a score of $S(I_{BG}) = 0.84$. 

\begin{table}[!tbh]
    \centering
    \begin{tabularx}{\linewidth}{|l|X|}
    \hline
    \textbf{Name} &   Amitabh Bachchan \\
    \hline
    \textbf{POB} & Allahabad \\
    \hline
    \textbf{DOB} & 1942-10-11 \\
    \hline
    \textbf{Education} & Delhi University \\
    \hline
    \textbf{Career} & Actor,  Artist, Assistant, Producer  \\
    \hline
    \textbf{Awards} & Padma Vibhushan, Padma Bhushan, Padma Shri, Government of India\\
    \hline
    \end{tabularx}
    \caption{An example of an Infobox generated using \texttt{BioGen}.}
    \label{tab:infobox_example}
    \vspace{-0.4cm}
\end{table}

\section{Conclusion and Future Work}
In this work, we proposed a system that generates a biography of a person, given the name and reference documents as input. However, we aim to build a system that just takes the name as an input and generates a biography by extracting information from the web. We would also like to enhance our system by incorporating coreference resolution so that it takes care of identifying the sentences related to the entity we are interested in. Right now, the system works well for extractive summarization, but we feel that adding in the abstractive form would help in creating better biographies.
One of the other enhancements to check would be neural network based models for the classification and sentence generation task.

\bibliographystyle{ACM-Reference-Format}
\bibliography{sigproc}
\end{document}